\documentclass[twocolumn,showpacs,preprintnumbers,amsmath,amssymb,aps]{revtex4}
\input epsf
\usepackage{graphicx}
\begin{document}

\title{Creating Artificial Ice  States Using Vortices in Nanostructured Superconductors} 
\author{A. Lib{\' a}l, C.J. Olson Reichhardt,
and C. Reichhardt} 
\affiliation{ 
Theoretical Division, Los Alamos National Laboratory, Los Alamos, New Mexico 87545
}

\date{\today}
\begin{abstract}
We demonstrate 
that it is possible to realize  
vortex ice states that are analogous to square and kagom{\' e} ice. 
With numerical simulations, we show that
the system can be brought into a state 
that obeys either global or local ice rules by applying an external current
according to an annealing protocol.  We explore the breakdown of the ice rules
due to disorder in the nanostructure array and show that 
in square ice, topological defects appear along grain 
boundaries, while in kagom{\' e} ice, individual defects appear. 
We argue that the vortex system offers significant advantages over 
other artificial ice systems. 
\end{abstract}
\maketitle

\vskip2pc

Geometric frustration occurs when a system is constrained by 
geometry in such a way that the 
pairwise interaction energy cannot be simultaneously 
minimized for all constituents,
and appears
in water ice \cite{Pauling}, spin systems 
\cite{Anderson,Moss,Ramirez},
and a variety of other systems in both 
physics \cite{Lubenky} and biology \cite{Lowen}. 
A specific example of frustration occurs in the
classical spin ice system where the constituents of the system are 
magnetic spins 
on a grid of 
corner-sharing tetrahedra.  The 
spins are constrained to point along the lines connecting the middle 
points of the tetrahedra 
\cite{Moss,Ramirez} 
and pairs of spins can 
minimize their energy by adopting a head-to-tail configuration. 
It is not, however, possible
for the four spins on a 
tetrahedron to simultaneously 
satisfy each of the six pairwise interactions in a 
head-to-tail fashion; 
the best the system can do is to 
satisfy four 
interactions out of six, 
leaving two pairs in a head-to-head or tail-to-tail configuration. 
As a result, in the ground state configuration each 
tetrahedron obeys the so-called ``ice rule'' of a two-in two-out 
configuration with two spins pointing toward the center of the
tetrahedron and two spins pointing away from it.  
Defects appear in the form of magnetic monopoles \cite{Castelnovo}.

Recently, there has been growing interest in creating model systems 
that exhibit spin ice behavior 
\cite{Wang,Nisoli,Ke,Moller,Tanaka,Qi,Libal}       
and that allow the individual constituents to be imaged directly, unlike
molecular or atomic ices.
For example, Wang {\it et al.} \cite {Wang}
created artificial square ice using
single-domain rectangular ferromagnetic islands 
arranged in a square lattice such that four islands meet 
at every vertex point.  
They found that as the inter-island interaction increased, 
the system preferentially formed ice-rule-obeying vertices, but
it did not reproduce the known ground state of two-dimensional (2D)
spin ice, where the two ``in'' 
magnetic moments are on opposite sides of the
vertex.
This could be due to the relative weakness of the 
magnetic interactions.
It has recently been shown that 
certain dynamical annealing protocols permit the system 
to approach the ground state more closely \cite{Nisoli,Ke}.
Similar studies have been performed 
for a 
2D kagom{\' e} ice system \cite{Tanaka,Qi}
where the local ice-rules were obeyed 
and defects such as three-in or three-out were absent \cite{Qi}.     
In the colloidal artificial ice system of Ref.~\cite{Libal},
the local dynamics can be accessed easily via video microscopy;
however, the ice arrays in this system 
are limited to relatively small sizes in experiment.

Here we propose that a 
particularly promising artificial ice system could be created using
vortices in superconductors with appropriately designed nanostructured 
arrays of artificial pinning sites. 
There has been extensive experimental work 
showing that a rich variety of different pinning array geometries 
can be fabricated 
\cite{Baert,Schuller,Harada,Field,Karapetov,Fedor}, and
various types of experimental techniques exist for directly imaging vortices 
in these arrays 
\cite{Harada,Field,Karapetov,Ir}.    
The vortex system has several advantages 
over other artificial ice systems. 
The vortex-vortex interaction strength is large, permitting the ground state
to be reached much more readily than in the nanomagnetic systems.
An applied external current permits the straightforward realization of
different dynamical annealing protocols. 
New types of defects can be studied 
by merely increasing or decreasing the magnetic field to 
create vacancies or interstitials that locally break the ice rules, 
while transport properties and critical currents 
can be measured which are not accessible in the other systems. 

To form square vortex ice, we propose using
an arrangement of elongated double-well pinning 
sites. 
Nonsuperconducting islands with the double-hump shape illustrated in
Fig.~1(a) placed within a superconducting layer 
have a pair of potential minima 
at the highest points of the island where the superconducting layer is
the shallowest.
A single vortex trapped over each island will sit at one
of the two minima, depending on the interactions with nearby vortices.
By changing the arrangement of the islands, different types of ice can
be created.  For square ice, shown in Fig.~1(a), four islands come together at
each vertex and the state of each island 
is defined as ``in'' if the vortex sits close to the vertex and ``out'' 
otherwise.  We define $n_{\rm in}$ as the number of ``in''
vortices at a 
vertex.
In Fig.~1(a), the vortices have formed
an 
$n_{\rm in}=2$ ice-rule-obeying ground state configuration. 
Figure 1(b) 
shows a kagom{\' e} spin ice arrangement with three islands 
surrounding each vertex.
In this case, the lowest energy state has $n_{\rm in}=1$ or $n_{\rm in}=2$
at each vertex, but there is no overall ordering into 
a unique ground state. 

\begin{figure}
\includegraphics[width=\columnwidth]{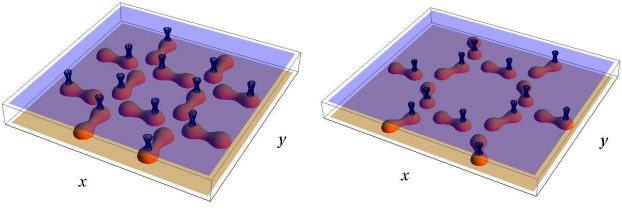}
\caption{Schematic of the nanostructured pinning site configurations producing
ice states.  Double-lobed objects: pins; 
open mesh objects: 
vortices. 
a) Square ice ground state.
b) One possible biased ground state of the kagom{\' e} ice system.}
\label{fig_schematics}
\end{figure}

To study the vortex ice, 
we perform numerical simulations of a 
2D sample with
periodic boundaries
containing $N_p$ elongated pinning sites in the square or kagom{\' e}
configurations illustrated in Fig.~1 and $N_v=N_p$ vortices.
A vortex $i$ at position ${\bf R}_i$ 
obeys the following overdamped equation of motion:
$\eta (d{\bf R}_{i}/dt) = {\bf f}_{i}^{vv}  
+ {\bf f}^{s}_{i} + {\bf f}^{d} + {\bf f}^T_i.$
The damping constant $\eta=\phi_0^2d/2\pi\xi^2\rho_N$,
where  
$\phi_0=h/2e$ is the flux quantum, $\xi$ is the 
superconducting coherence length, $\rho_N$ is the normal state 
resistivity of the material, and $d$ is the thickness of the superconducting
crystal. 
The vortex-vortex interaction force 
is given by
${\bf f}_{i}^{vv} = \sum^{N_{v}}_{j\ne i}f_{0}
K_{1}(R_{ij}/\lambda){\bf {\hat R}}_{ij},$
where $K_{1}$ is the modified Bessel function appropriate for stiff 
three-dimensional vortex lines, $\lambda$ is the London penetration depth, 
$f_{0}=\phi_0^2/(2\pi\mu_0\lambda^3)$, $R_{ij}=|{\bf R}_i-{\bf R}_j|$,
and ${\bf {\hat R}}_{ij}=({\bf R}_i-{\bf R}_j )/R_{ij}$.
The substrate force ${\bf f}^{s}_{i}$ arises from the elongated pins, 
${\bf f}^{s}_{ik}=
\sum_{k}^{N_p}f_0(f_p/r_p)R_{ik}^\pm \Theta(r_p-R_{ik}^\pm){\bf \hat{R}}_{ik}^\pm
+f_0(f_p/r_p)R_{ik}^\perp \Theta(r_p-R_{ik}^\perp){\bf \hat{R}}_{ik}^\perp
+f_0(f_b/l)(1-R_{ik}^\parallel) \Theta(l-R_{ik}^\parallel)
{\bf \hat{R}}_{ik}^\parallel$.
Here
$R_{ik}^\pm=|{\bf R}_i-{\bf R}_k^p \pm l{\bf {\hat p}}^k_\parallel|$,
$R_{ik}^{\perp,\parallel}=
|({\bf R}_i - {\bf R}_k^p) \cdot {\bf {\hat p}}^k_{\perp,\parallel}|$,
${\bf R}_k^p$
is the position of pin $k$, 
and ${\bf \hat{p}}^k_\parallel$ (${\bf \hat{p}}^k_\perp$) is a unit
vector parallel (perpendicular) to the axis of pin $k$.
Each vortex is constrained to stay within a pin composed of two 
half-parabolic wells of 
radius $r_p=0.4\lambda$ 
separated by an 
elongated region of length $2l$ which confines the 
vortex 
perpendicular to the pin axis and has a repulsive potential or barrier of 
strength $f_b$ parallel to the axis which pushes the vortex out of the middle 
of the pin into one of the ends.
We take 
$l=2/3\lambda$ or $5/6\lambda$ and vary
the lattice constant $a$ of the pinning array
between $a=2.0\lambda$ and $8.0\lambda$.
The driving force ${\bf f}_d$ represents the Lorentz force from an applied
current.
The thermal force ${\bf f}^T_i$ comes from thermal Langevin kicks and is
set to zero except during the annealing of the kagom{\' e} ice.

{\it Square Ice -}
We prepare the square ice 
system using a dynamical annealing procedure inspired by
the nanomagnetic ice results of Refs.~\cite{Nisoli,Ke}. 
In our simulations, we place one vortex in each pin at a random position
and then use a protocol of a rotating
in-plane applied current with decreasing amplitude, 
${\bf f}^{d}= A_{ac}(t)(\cos(2\pi t/T_r){\bf \hat{x}} 
+ \sin(2\pi t/T_r){\bf \hat{y}})
$,
where $T_r=1000$ simulation time steps, 
$A_{ac}(t)=\pm(A_0-\delta A \lfloor t/\delta t\rfloor)$,
$A_0=2.0f_0$, $\delta t=10000$ simulation time steps, 
and $\delta A=0.01f_0$.
The force direction is reversed
each time the magnitude of the force is decreased.
We measure the number of vertices of each type that appear after 
completing the
dynamical annealing. 
For the kagom{\' e} ice system, we 
obtain the vortex configurations from standard thermal 
simulated annealing.

To determine how effectively the dynamical
annealing protocol brings the square ice system to the 
ground state, 
we introduce disorder to the system by replacing the delta-function 
distributed barriers $f_b$ at the center of each pinning site with 
barriers of normally distributed strength, where the mean strength is
$f_b$ and the width of the distribution is $\sigma$.
In Fig.~2 
we illustrate the vertices that have reached the ground state configuration
of 
$n_{\rm in}=2$ in a square ice sample with
$a=2.5\lambda$, $l=5/6\lambda$ and $f_b=0.25f_0$
for differing disorder widths $\sigma$. 
The dots represent vertices in the ice-rule obeying ground state, 
while the closed black circles 
indicate higher energy vertices 
that we term ice-rule defects $D_{\rm I}$ since they 
still obey the 
$n_{\rm in}=2$ ice rule
but have the two ``in'' vortices adjacent to one another.
The open circles mark
the highest energy vertices
that we term non-ice-rule defects $D_{\rm NI}$ 
since they do not obey the 
$n_{\rm in}=2$ ice
rule but have, for example, 
$n_{\rm in}=3$ or $n_{\rm in}=0$.
For  $\sigma < 0.1$, 
the system can reach the ordered ground state as shown in Fig.~2(a).   
As the central barriers of the pins become more nonuniform with increasing
$\sigma$, some pinning centers act as nucleation sites for grain boundaries,
as illustrated in Figs.~2(b,c) for $\sigma=0.1$ and $\sigma=0.5$.
In general, we find that for $0.1 < \sigma < 0.7$, all of the defected
vertices
form closed loop grain boundaries and 
the ratio of 
$D_{\rm I}$ to $D_{\rm NI}$ is 1:1 due to geometric constraints. 
For $\sigma \ge 0.7$, Fig.~2(d) shows that a proliferation of 
$D_{\rm NI}$ occurs so that
the $D_{\rm NI}$ outnumber the $D_{\rm I}$.
The grain boundary loops interact and wind around the sample, making it
difficult to determine the relation between $\sigma$ and the grain boundary
length.
We find that individual 
$D_{\rm NI}$
can appear outside of grain boundaries, while 
$D_{\rm I}$ always
remain confined to grain boundaries,
suggesting that there could be a 
disorder-induced phase transition when the 
$D_{\rm NI}$
proliferate.
We also find that 
doubly occupied pinning sites with two vortices each
can act as grain boundary nucleation sites,
as illustrated in the inset of Fig.~4(b).

\begin{figure}
\includegraphics[width=\columnwidth]{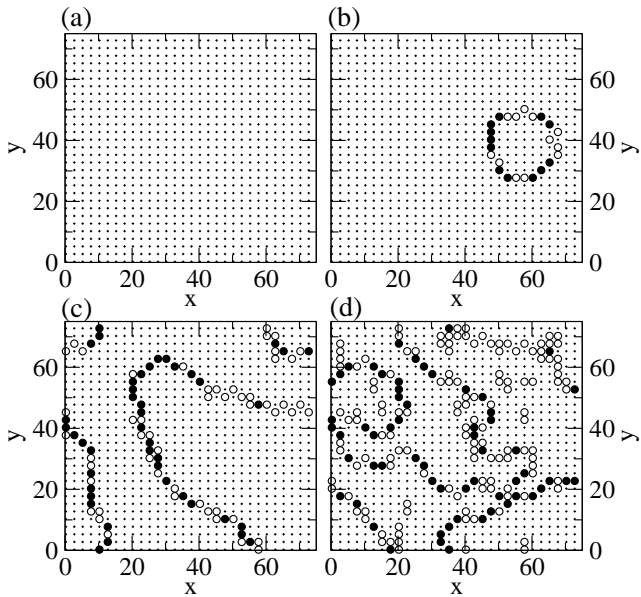}
\caption{Grain boundary images in square ice samples with 
$a=2.5\lambda$, $l=5/6\lambda$, and $f_b=0.25$ for increasing disorder width
$\sigma$.
Dots: ground state $n_{\rm in}=2$
ice-rule obeying
vertices; 
filled black circles: ice-rule defects $D_{\rm I}$; 
white circles: non-ice-rule defects $D_{\rm NI}$.
(a) $\sigma=0$. (b) $\sigma=0.1$. (c) $\sigma=0.5$.
(d) $\sigma=1.0$.
}
\label{fig_grainboundaries}
\end{figure}

In Figure 3(a), we plot the percentage of vertices $P_{GS}$ 
that are in the ice-rule obeying ground state 
as a function of time during the dynamical annealing procedure
in a sample with $a=2.5\lambda$, $l=5/6\lambda$, $f_b=0.25f_0$, 
and different values of $\sigma$.
At early times, when $|A_{ac}|$ is close to $A_0$, all of the vortices
follow the drive and switch back and forth inside the pinning sites.  As
$|A_{ac}|$ decreases, a transition occurs when the vortices cease to follow
the driving direction and become locked into one position in the pinning
site.  For $\sigma=0$, this locking transition is 
relatively sharp and occurs 
at $|A_{ac}|\approx 0.82f_0$.  Nonzero values of $\sigma$ 
broaden the transition significantly and cause some vertices to lock into
the ground state at much earlier times; at the same time, complete locking of
all vertices into the ground state can no longer be achieved within the
finite time of the dynamical annealing process.
We quantify the broadening of the transition with increasing $\sigma$ by
fitting the curves in Fig.~3(a) to the form
$P_{GS}(t)=1-\exp(t/\tau)$.
Figure 3(b) shows the fitted 
relaxation time $\tau$ as a function of $\sigma$ and indicates
the occurrence of an increasingly slow locking process as the disorder
width increases.
The dependence of $P_{GS}$ on both $a$ and $\sigma$ is
summarized in Fig.~3(d) for a system with $f_b=1.0$ and $l=2/3\lambda$. 
Here, $P_{GS}$ decreases both with increasing $\sigma$ and with
increasing $a$ as the relative strength of the 
vortex-vortex interactions decreases.

Depending on the system parameters, it is not always necessary to perform
a dynamical annealing procedure in order to reach the ground state.
To demonstrate this, we prepare the sample in a random state and then
apply a fixed amplitude rotating drive,
${\bf f}^d=\tilde{A}(\cos(2\pi t/T_r){\bf {\hat x}}+\sin(2\pi t/T_r){\bf {\hat y}})$,
with $\tilde{A}=0.01f_0$ and $T_r=1000$ simulation time steps,
for $2\times 10^6$ simulation time steps.
When the central barrier in the pin $f_b$ is weak, 
the system can reach the ordered ground state under the weak
external shaking.  For larger $f_b$, the system cannot 
reach the ordered
ground state without dynamical annealing.
This is shown in Fig.~3(c), 
where we plot the final $P_{GS}$ at the end of the simulation time
versus $f_{b}$ for samples with $\sigma = 0.01$ and 
varied pinning lattice constant $a=2.0\lambda$, $2.5\lambda$, and 
$3.0\lambda$.
For large $f_b$, the sample is immediately frozen into the disordered
initial configuration, and $P_{GS}\approx 0.125$, consistent with the
value expected in a completely random sample.
As $f_{b}$ is lowered, a spontaneous 
rearrangement into a partially ordered state 
becomes possible and $P_{GS}>0.125$.
The value of $f_b$ at which the spontaneous ordering 
appears
increases with decreasing $a$, indicating that as the vortex-vortex
interactions grow stronger in the denser pinning arrays, the ordered
ground state is much more energetically favored.

\begin{figure}
\includegraphics[width=\columnwidth]{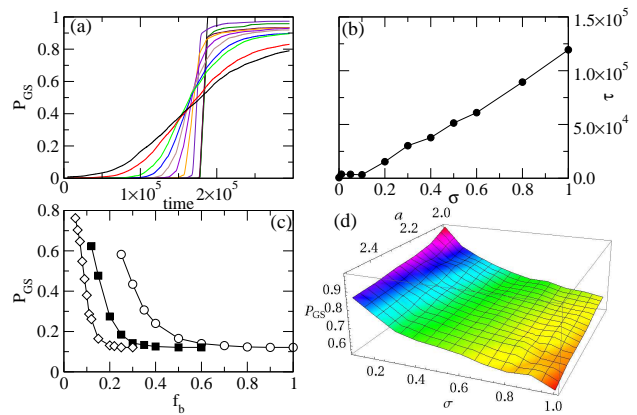}
\caption{(a) Percentage $P_{GS}$ of ice-rule obeying ground state vertices
vs time during the dynamical annealing process for different
disorder widths $\sigma$.  From upper right to lower right,
$\sigma=0$, 0.01, 0.05, 0.1, 0.2, 0.3, 0.4, 0.5, 0.6, 0.8,
and 1.0.   Here, $a=2.5\lambda$, $l=5/6\lambda$, and $f_b=0.25$.
(b) Relaxation time $\tau$ vs $\sigma$ for the
same system.
(c) 
Final value of $P_{GS}$ vs $f_b$ in samples 
subjected to a small shaking field
with no dynamical annealing. 
Here $l=5/6\lambda$, $\sigma=0.1$,
and $a=2.0\lambda$ (open circles), $2.5\lambda$ (filled squares), 
and $3.0\lambda$ (open diamonds).
(d) $P_{GS}$ 
vs 
$\sigma$ and $a$ in a sample with 
$f_b=1.0$ and $l=2/3\lambda$.
}
\label{fig_plots}
\end{figure}

\begin{figure}
\includegraphics[width=\columnwidth]{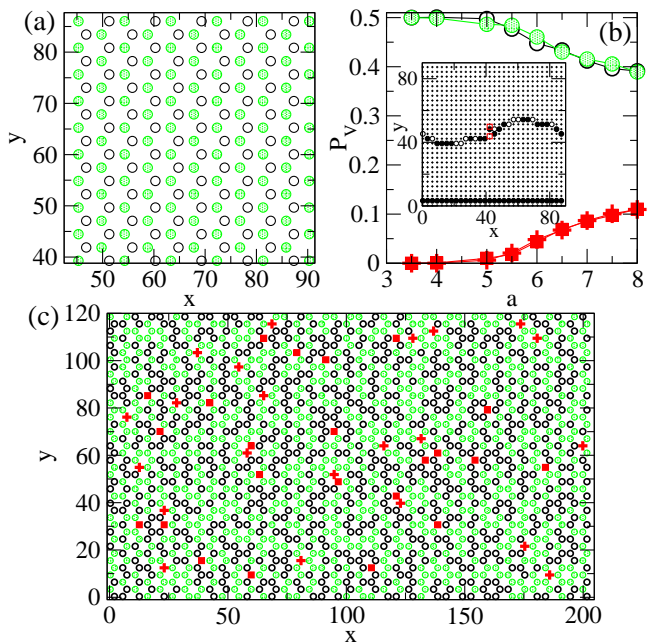}
\caption{
(a) Ordered biased ground state in a sample with kagom{\' e} pinning, 
$f_b=1.0$, $l=2/3\lambda$, and
$a=3\lambda$.  Open circles: $n_{\rm in}=1$ vertices; 
shaded circles: 
$n_{\rm in}=2$ vertices.
(b) Percentage $P_V$ of each vertex type 
vs $a$.
Crosses: $n_{\rm in}=0$; 
open circles: $n_{\rm in}=1$; 
shaded circles: $n_{\rm in}=2$; 
filled squares: $n_{\rm in}=3$. 
Inset: grain boundary image in square ice sample with two doubly occupied
pins (open squares) with the same symbols as in Fig. 2.
(c) Vertex configuration after thermal annealing in a sample with 
$a=3.5\lambda$, $l=2/3$, and $f_b=1.0$.  Symbols are the same as in the
main panel of (b).
}
\label{fig_hexa}
\end{figure}

{\it Kagom{\' e} ice - } 
The kagom{\' e} lattice illustrated in Fig.~1(b) has a distinct set of ice
rules from the square lattice. 
High energy vertices with 
$n_{\rm in}=0$ or 3 are avoided in favor of the kagom{\' e}-ice-rule obeying
vertices with 
$n_{\rm in}=1$ or 2.
This system 
can form a 
non-unique ordered ground state, but only in
the presence of an external biasing field.
In Fig.~4(a) we show 
one possible biased ordered ground state for a kagom{\' e}
lattice with 
$f_b=1.0$ 
and $\sigma=0$ obtained by applying a constant drive 
${\bf f}^d=0.01f_0({\bf {\hat x}}+{\bf {\hat y}})$ along a lattice symmetry
direction while performing simulated annealing.
In the absence of the biasing force, some high energy defect vertices
which take the form of monopoles
appear in the system and there is no overall order, as illustrated in
Fig.~4(c).
We find that the kagom{\' e} ice
is more robust against the effects of disorder 
than the square ice,  
in
agreement with experimental findings for nanomagnetic 
kagom{\' e} ice \cite{Qi}. 
The 
defect patterns are distinct from the square ice 
since no grain boundary state
forms for the kagom{\' e} ice due to the lack of an ordered ground state.       
Unlike the bipartite square lattice, the nonbipartite 
kagom{\' e} lattice is not topologically constrained, making
our system more closely resemble the ice state
studied in Ref.~\cite{Wills} than that
considered in Ref.~\cite{Sondhi}.
Although Fig.~4(c) shows that there is some tendency 
for the defected vertices to form pairs,
there are no extended defect patterns of the type
seen in Fig.~2. 
Since the ice rules in this system are enforced by the 
vortex-vortex interaction energies,
we can weaken the enforcement of the ice rules by increasing the
spacing $a$ between pinning sites.
Figure 4(b) shows that as $a$ increases, the system passes from a limit
in which only kagom{\' e}-ice-rule obeying vortices appear for $a\le 4\lambda$
to a limit $a \ge 8\lambda$ where the vertices assume a completely random
arrangement.  In the random limit we expect to find 
each of the two 
defect vertex types 
with
probability $1/8$ and each of the two kagom{\' e}-ice-rule obeying vertices
with probability $3/8$.

There are other arrays that would obey
ice-rule type constraints; however, the simplest cases for 
2D are the square and kagom{\' e} arrays.
Previous studies of superconducting wire networks arranged in kagom{\' e}
configurations found geometrical frustration which produced disordered
ground states
\cite{Field2};
however, such a system does not specifically have ice-rule obeying states.  
The artificial ice vortex system 
proposed here can be used 
to study the effect of ice-rule and non-ice-rule
configurations on transport and magnetization properties, and it would 
also be possible to examine higher matching fields to see whether new types of
ordered or disordered states appear.

In summary, 
we propose that square and 
kagom{\' e} vortex ice can be realized in nanostructured superconductors.  
By using 
an annealing protocol 
of a rotating externally applied current, the system can 
reach or approach
the
square ice ground state.
In the presence of quenched disorder, 
defects 
appear in 
an ordered ground state background.
For moderate disorder in the square ice system, all of the defects are
bound to grain boundaries, 
while for strong disorder, individual high energy vertices proliferate. 
For kagom{\' e} ice, we find no grain boundary phase in the presence
of disorder.
We predict that if the barrier for vortex motion 
across the center of each artificial pinning site
is 
weak, 
the system will spontaneously organize 
into a partially ordered state even without use of an annealing protocol.
This system could have interesting transport and memory 
effects which may manifest themselves as changes in the
critical current, an effect which cannot be accessed readily
in other artificial ice systems.  

We thank C. Nisoli for a
useful discussion.
This work was carried out under the 
NNSA of the U.S.~DoE
at LANL under Contract No.~DE-AC52-06NA25396.

\end{document}